\documentclass[fleqn,twoside]{article}
\usepackage[headings]{espcrc2}

\readRCS
$Id: espcrc2.tex,v 1.2 2004/02/24 11:22:11 spepping Exp $
\usepackage{graphicx, balance}
\usepackage[figuresright]{rotating}

\usepackage{epstopdf} 
\usepackage{epsfig}

\usepackage{graphicx}
\usepackage{caption}
\usepackage{subcaption}
\usepackage{cases}

\newcommand{\AmS}{{\protect\the\textfont2
  A\kern-.1667em\lower.5ex\hbox{M}\kern-.125emS}}

\title{\textbf{Fisher-Yates Chaotic Shuffling Based Image Encryption}}

\author{Swaleha Saeed\address[MCSD]{Department of Computer Engineering, ZH College of Engineering and Technology, Aligarh Muslim University, Aligarh 202 002, India, Contact: \{swalehasaeed, saroshumar, atharali\}@zhcet.ac.in}%
        ,
        M Sarosh Umar\addressmark,
		M Athar Ali\addressmark,
        Musheer Ahmad\address{Department of Computer Engineering,
Faculty of Engineering and Technology,
Jamia Millia Islamia, New Delhi 110 025, India, Contact: musheer.cse@gmail.com}
        }

\runtitle{Fisher-Yates Chaotic Shuffling Based Image Encryption}
\runauthor{Swaleha Saeed, et al.,}

\begin{document}
\begin{abstract}

{\bf Abstract :} 
In Present era, information security is of utmost concern and encryption is one of the alternatives to ensure security. Chaos based cryptography has brought a secure and efficient way to meet the challenges of secure multimedia transmission over the networks.  In this paper, we have proposed a secure Grayscale image encryption methodology in wavelet domain.  The proposed algorithm performs shuffling followed by encryption using states of chaotic map in a secure manner. Firstly, the image is transformed from spatial domain to wavelet domain by the Haar wavelet. Subsequently, Fisher Yates chaotic shuffling technique is employed to shuffle the image in wavelet domain to confuse the relationship between plain image and cipher image. A key dependent piece-wise linear chaotic map is used to generate chaos for the chaotic shuffling. Further, the resultant shuffled approximate coefficients are chaotically modulated. To enhance the statistical characteristics from cryptographic point of view, the shuffled image is self keyed diffused and mixing operation is carried out using keystream extracted from one-dimensional chaotic map and the plain-image. The proposed algorithm is tested over some standard image dataset. The results of several experimental, statistical and sensitivity analyses proved that the algorithm provides an efficient and secure method to achieve trusted gray scale image encryption.  
\end{abstract}

% typeset front matter (including abstract)
\maketitle

\section{INTRODUCTION}
Nowadays, public networks are not suitable for the direct transmission of confidential messages because of the rapidly rising problem of security threats. As a result, security is still an open challenge in spite of the tremendous advancement in internet technologies. This is an important challenge in the areas where reliable, secure, fast and robust transmission of information is the major requirement especially in the field of military and medical systems. To deal with the problem of secure transmission of information over the networks, numerous encryption algorithms have been proposed based on different methodologies and ideas [1][2][3].\\

Traditional ciphers RSA, DES, AES, can be used to encrypt image, but these are not ideal for two reasons [4]. First, since image size is generally much greater than text. This results in conventional ciphers taking much more time to encrypt images. Second, image data has high correlation among adjacent pixels. Consequently, it is rather difficult for these ciphers to shuffle and diffuse image data effectively. Chaos-based cryptosystems usually have higher speeds and lower costs. In this regard, chaos based encryption techniques have demonstrated exceptionally good behavior because these technique have faster speed, reasonable computation overheads without compromising the security. Moreover, chaotic systems have many important properties such as the sensitivity to initial conditions and system parameters, pseudorandom property, non-periodicity and topological transitivity [1]. These optimistic features make the chaotic algorithm ideal for image encryption. A number of chaos- based techniques for multimedia encryption have been proposed in recent times [5]. \\

The proposed image encryption algorithm comprises two phases - shuffling of plain image; and self keyed diffusion followed by mixing operation. The primary goal of shuffling is to transform a meaningful image into a meaningless, disordered version so as to obscure real meaning of image. Further, it also reduces the correlation coefficient between neighbour pixels and produces data diffusion.  However, shuffling in spatial domain changes the position of pixels in plain image but the statistical information is left intact even after shuffling. So, it is an insecure approach to perform shuffling in spatial domain as the attacker can utilize the characteristics of shuffled image and can recover the plain-image [6].\\

In this paper, a secure key dependent shuffling is performed in wavelet domain by exploiting the feature of classical Fisher-Yates shuffle technique. The shuffled approximate wavelet coefficients are further chaotically modulated to achieve desired level of confusion and diffusion. The resultant shuffled image is self keyed diffused and then mixing operation is performed using keystream extracted from one-dimensional chaotic map and the plain-image.\\

    The structure of rest of this paper is organized as follows: the preliminaries are described in subsequent Section 2. The proposed encryption method is discussed in Section 3. The results of simulation are analysed in Section 4, while the conclusions of the work are summarized in Section 5.

\section{PRELIMINARIES}

\subsection{Piece-Wise Linear Chaotic Map}
Piece-Wise Linear Chaotic maps (PWLCM) are the simplest kind of chaotic map from realization point of view for both hardware and software. They possess perfect dynamical properties, hence widely used in digital chaotic ciphers.  The general form to define piecewise linear chaotic map is given in (1).\\
\begin{numcases}{x_{n+1}=}
 \frac{x_n}{m} & for $0 < x_n < m $
\\ \nonumber
\frac{1-x_n}{1-m} & for $ m\leq x_n < 1 $
\end{numcases}

where $x_n$ and {\it m} are state variable and system parameter, respectively. The value of {\it m} should be restricted in the range (0,1) for generation of chaotic random sequence. The PWLCM map behaves chaotically for 0 $<$ {\it m} $<$ 1  and a random sequence is generated on iterations.

\subsection{Fisher-Yates Shuffle}
The Fisher-Yates Shuffling technique in its original form was suggested by Ronald Fisher and Frank Yates, also known as the Knuth shuffle in 1938. Basically, the algorithm is used for generating a random permutation of a finite linear array. The technique is further modernized by the Richard Durstenfeld [7] and popularized by Donald E. Knuth in his pioneer book \textit{The Art of Computer Programming} [8]. To shuffle an array {\it S} of n elements (indices 1...n), perform the following step-
\\ \\
\textit{for i $\leftarrow$ n to 1 do\\
j $\leftarrow$ random integer with 1 $\leq$ j $\leq$ i\\
exchange(S[j], S[i])\\
end}
\\ \\
The most promising property of Fisher-Yates shuffle is that it produce an unbiased result that is, every permutation of the array is equally likely. The modern version is an in-place shuffle, means that given a pre-initialized array, it shuffles the elements of the array in place rather than producing a shuffled copy of the array. So, it is an efficient algorithm, requiring only time proportional to the number of elements being shuffled and also it does not need any additional storage space. However, the shuffling effect of algorithm depends only on the quality of random generation of indices. So, it can be improved further by incorporating a piece-wise linear chaotic map as source to generate random indices and applying the iterations of algorithm on previously shuffled linear array.
 
\subsection{Discrete Wavelet Transform}
The Discrete Wavelet Transform DWT is a linear transformation that provides a compact representation of signal’s frequency with strong spatial support. DWT can decompose a signal into its frequency sub-bands at different scales from which it can be reconstructed back to get the original image. The proposed scheme employs \textit{Haar} filter for wavelet decomposition of the plain-image data to transform plain image from spatial domain to wavelet domain. The Haar wavelet is simplest wavelet transform which is discontinuous and resembles a step function. For a given function f, the Haar wavelet transform can be defined as (2). \\
\begin{equation}
 \left(
 \begin{array}{c} f \rightarrow (a_L \mid d_L)\\ a_L = (a_1,a_2,........,a_{N/2})\\
d_L = (d_1,d_2,........,d_{N/2})\end{array}
 \right) 
\end{equation}
where {\it L} is the decomposition level, {\it a} is the approximation sub-band and {\it d} is the detail sub-band. As shown in Figure 1 the Barbara image is decomposed at level 1 which provides the four sub-bands: approximate coefficients sub-band i.e. LL sub-band and detailed coefficient sub-bands i.e. HL, LH, and HH sub-bands. The LL sub-band represents approximation of the original image while the other sub bands contain the missing details. The sub-bands LH, HL and HH highlight the horizontal (cH), vertical (cV) and diagonal (cD) coefficients respectively. Since, LL sub-band contains major portion of image information, it can be decomposed further at subsequent levels. \\

Using inverse discrete wavelet transform, the sub-bands can be combined to get back the image. The inverse of the Haar wavelet transform is computed in the reverse order by (3).
\begin{eqnarray}
f & = & \frac{a_1 - d_1}{\sqrt{2}}, \frac{a_1 + d_1}{\sqrt{2}},.......
 \\ \nonumber
& & ....., \frac{a_{N/2} - d_{N/2}}{\sqrt{2}}, \frac{a_{N/2} + d_{N/2}}{\sqrt{2}}   
\end{eqnarray}

Wavelet Transform is widely used in areas of image processing such as image and video processing, feature detection and recognition, image denoising and face recognition etc.
\begin{figure}
\centering
\resizebox{6cm}{6cm}{\includegraphics{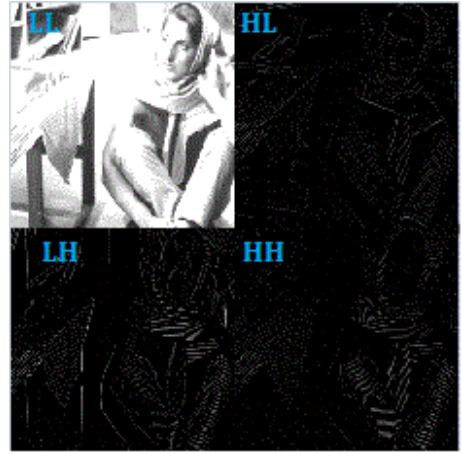}}
\centering
\caption{Wavelet decomposition of Barbara image at level 1}
\end{figure}

\section{PROPOSED METHOD}
The proposed encryption methodology exploits the features of one dimensional piecewise linear chaotic map, Haar wavelet transform and Fisher-Yates shuffle technique. Firstly, the image is transformed into wavelet domain using Haar wavelet transform. Then, Fisher-Yates shuffle is employed and indices of shuffle are randomly generated through PWLCM. Further, the shuffled wavelet approximate coefficients are chaotically modified to enhance the statistical characteristics of the image. Inverse Discrete Wavelet Transform (IDWT) is applied to get the final shuffled image and the resultant image is then self keyed diffused. To further enforce the security, mixing operation is performed in order to diffuse the gray values of image pixels using keystream extracted from chaotic map and the plain-image.
\\ \\First, the map described in (1) is iterated 1000 times to let the transient effect of map die-out and these generated values of state variable,  $x_1$ to  $x_{999}$, are then discarded. The proposed image encryption method is described as follows:
\begin{enumerate}
\item Let the size of original image is $M \times N$.
\item Apply DWT to decompose the image at level 1 which provides four wavelet coefficients sub-bands: {\it cA, cH, cV, cD} each of size $r \times c$.
\item Reshape the 2D sub-bands {\it cA, cH,cV and cD} to 1D arrays of size {\it p} (= r$\times$c).
\item Read $n_1$, $n_2$, $n_3$, $n_4$ ($n_1 < n_2 < n_3 < n_4$) and take $\xi$= 1.
\item Set {\it count} = 1.
\item Further iterate the map (1) and sample the chaotic state-variable {\it x}.
\item Extract a random number {\it m} $\in$ [1, k] from {\it x} as defined in (4).
\begin{equation}
m = \lfloor(x\times10^{10})\rfloor mod(k)+1
\end{equation}
where \textit{k = p $-$ count $+$ }1
\item Interchange the elements of 1D arrays of wavelet coefficients at positions {\it m} and {\it k} as follows:\\ \\
\textit{if ($\xi \leq n_1$)\\ 
$\lbrace$ \\
cD(k)$\leftrightarrow$cD(m) \\
cH(k)$\leftrightarrow$cH(m)\\ 
cV(k)$\leftrightarrow$cV(m)\\
cA(k)$\leftrightarrow$cA(m)}\\
$\rbrace$\\

\textit{if ($n_1  < \xi \leq n_2$)\\
$\lbrace$ \\
cH(k)$\leftrightarrow$cH(m)\\ 
cV(k)$\leftrightarrow$cV(m)\\
cA(k)$\leftrightarrow$cA(m)}\\
$\rbrace$\\

\textit{if ($n_2  < \xi \leq n_3$)\\
$\lbrace$ \\
cV(k)$\leftrightarrow$cV(m)\\
cA(k)$\leftrightarrow$cA(m)}\\
$\rbrace$\\

\textit{if ($n_3  < \xi \leq n_4$)\\
cA(k)$\leftrightarrow$cA(m)}\\

\item Extract key out of current map variable defined by (5).
\begin{equation}
key_v=\lfloor(x_{v+999}\times10^{15})\rfloor mod (256)
\end{equation}
where $v = (\xi-1)p + count $
\item Set $count = count + 1$, if $count < p$ goto Step 6.
\item Set $\xi = \xi + 1$, if $\xi \leq n_4$  goto Step 5. 
\item Further iterate the map (1) for \textit{p} times and store the values in array \textit{x(j)}, where \textit{j = 1,2,3,...,p}.
\item Generate a set $Y=\lbrace y(j) = x(j) \vert 1 \leq j \leq p\rbrace$ such that set \textit{y(j)} has 4-digits decimal precision.
\item Extract a binary sequence \textit{s(j)} based on the elements of \textit{Y}as specified in (6).
\begin{numcases}{s(j)=}
1 & if $y(j) > 0.5 $
\\ \nonumber
0 & {\it otherwise}
\end{numcases}

\item Change the sign of shuffled coefficients \textit{cA(j)}  wherever \textit{s(j)} is marked as 1.
\item Modulate the approximate coefficients \textit{cA} obtained above, chaotically in accordance with (7).
\begin{equation}
cA(j)' = \alpha \times y(j) \times cA(j)
\end{equation}           
where $y(j)\in Y$ and $\alpha$ in the range $(0<\alpha<1)$ is control parameter for modulation of approximate coefficients. Accordingly, the approximate coefficients can be demodulated in accordance to (8).
\begin{equation}
cA(j) = cA(j)' / (\alpha \times y(j))
\end{equation}

\item Reshape the four modified 1D arrays \textit{cA', cH, cV, cD} to 2D coefficient matrices of size $r \times c$.

\item Apply inverse wavelet transform to obtain the final shuffled image.
\item Reshape the 2D shuffled image to 1D array \textit{S} of size \textit{len}$(=M \times N)$.
\item Take $S_{len+1} = 170$ and perform keyed self diffusion as follows -
\begin{enumerate}
\item Repeat step (b) for $k = len$ to 2.
\item $S_{k-1} = S_{k-1} \oplus S_{k} \oplus S_{k+1}$.
\end{enumerate}
\item Perform mixing operation to compute pixel gray value of cipher image \textit{C}, using \textit{S} and \textit{key} as illustrated -
\begin{enumerate}
\item Take $C_0 = 85$ and $key_0 = 123$.
\item Repeat step (c) for $i = 1$ to $len$.
\item $C_i = S_i \oplus key_i \oplus \left[ key_{i-1} << 3 \right]  \oplus C_{i-1}$.
\\where $key_{i-1} << 3$ represent that $key_{i-1}$ is shifted by three times in left direction circularly.
\end{enumerate}
\item Reshape 1D array \textit{C} to 2D array to obtain final cipher image.
\end{enumerate} 
The block diagram of the proposed image encryption algorithm is shown in Figure 2. The plain-image can be recovered successfully by applying the proposed algorithm on cipher image in reverse order as the algorithm is symmetric in nature.

\begin{figure}
\centering
\resizebox{8cm}{15cm}{\includegraphics{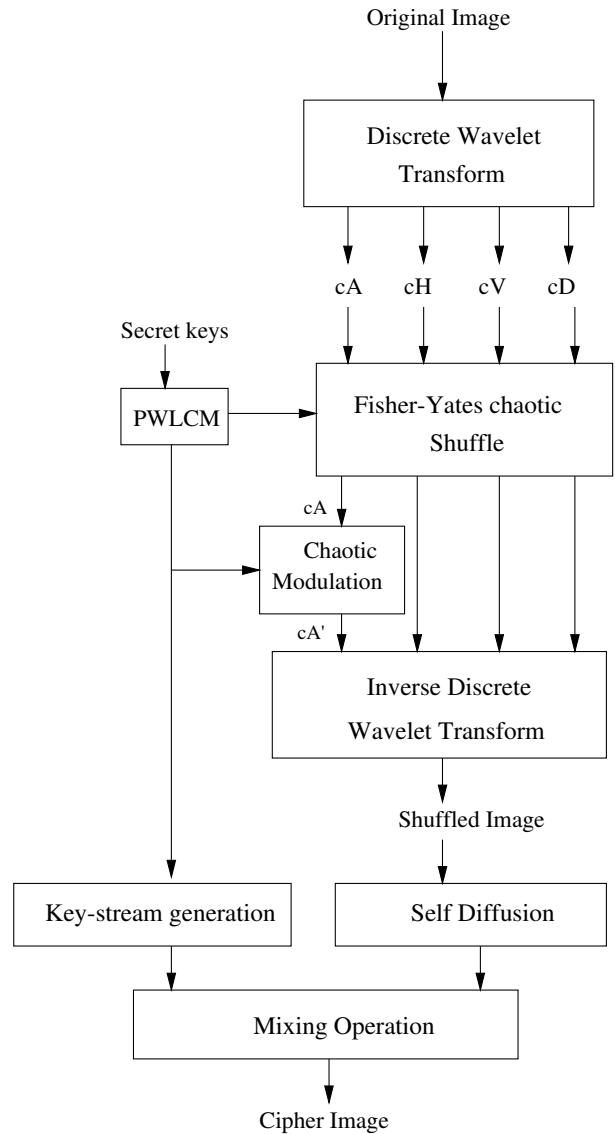}}
\caption{Proposed encryption algorithm}
\end{figure}

\section{SIMULATION ANALYSES}
The proposed encryption algorithm is implemented in MATLAB for computer simulations. The proposed encryption algorithm is experimented with various grayscale images like Baboon, Barbara, Lena, Boat, and Pepper of $256\times256$ size. The initial conditions and system parameters taken for simulations are: $x_0 = 0.123456, m = 0.489, n_1 = 1, n_2 = 2, n_3 = 3, n_4 = 4, \alpha = 0.2$. The secret key of proposed algorithm constitutes all these parameters.
\subsection{Histogram Analysis}
The image histogram represents the gray value distribution of pixels. The plain image of Barbara and cipher-image obtained by the proposed algorithm are shown in Figure 3(a) and 3(b) respectively. It is clear from Figure 3(a) and 3(b) that the cipher image is indistinguishable and completely visually disordered in comparison to the plain image. The decrypted image obtained using correct key is shown in Figure 3(c). Hence, it is clearly visible from Figure 3(c) that the proposed technique is successful in decrypting image from the encrypted version. \\

The histograms for original, cipher and decrypted image are shown in Figure 4(a), 4(b) and 4(c) respectively. The histogram of the encrypted image is fairly uniform and entirely different from the histogram of the original image. This proves that the cipher image doesn't provide any information regarding the distribution of gray values to the attacker. Hence, the proposed algorithm can resist any type of histogram based attacks. It can be concluded from Figure 4(c) the histogram of decrypted image is same as that for original image which proves no information loss in the encryption and decryption process. 
\begin{figure}
\centering
\begin{subfigure}[b]{0.3\textwidth}
\resizebox{5cm}{5cm}{\includegraphics{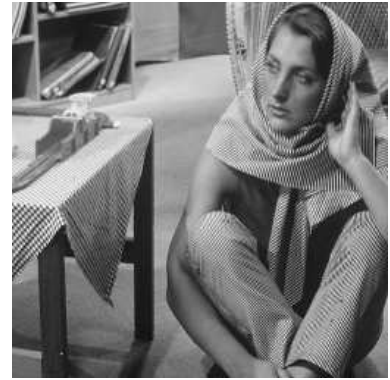}}
\caption{Original Image}
\end{subfigure}
\begin{subfigure}[b]{0.3\textwidth}
\resizebox{5cm}{5cm}{\includegraphics{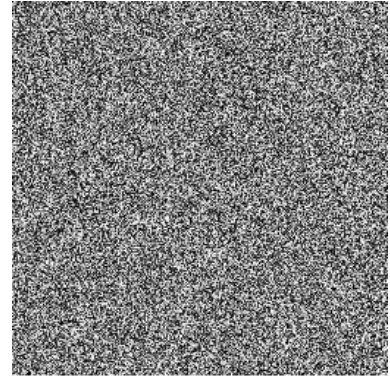}}
\caption{Cipher Image}
\end{subfigure}
\begin{subfigure}[b]{0.3\textwidth}
\resizebox{5cm}{5cm}{\includegraphics{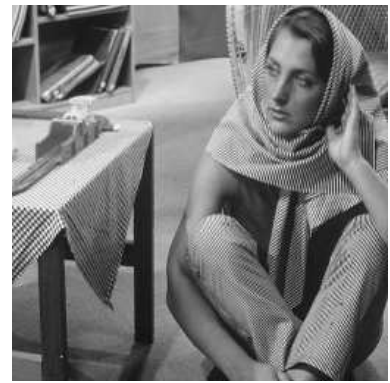}}
\caption{Decrypted Image}
\end{subfigure}
\caption{Encryption/Decryption result}
\end{figure}

\begin{figure}
\begin{subfigure}[b]{0.3\textwidth}
\resizebox{7cm}{6cm}{\includegraphics{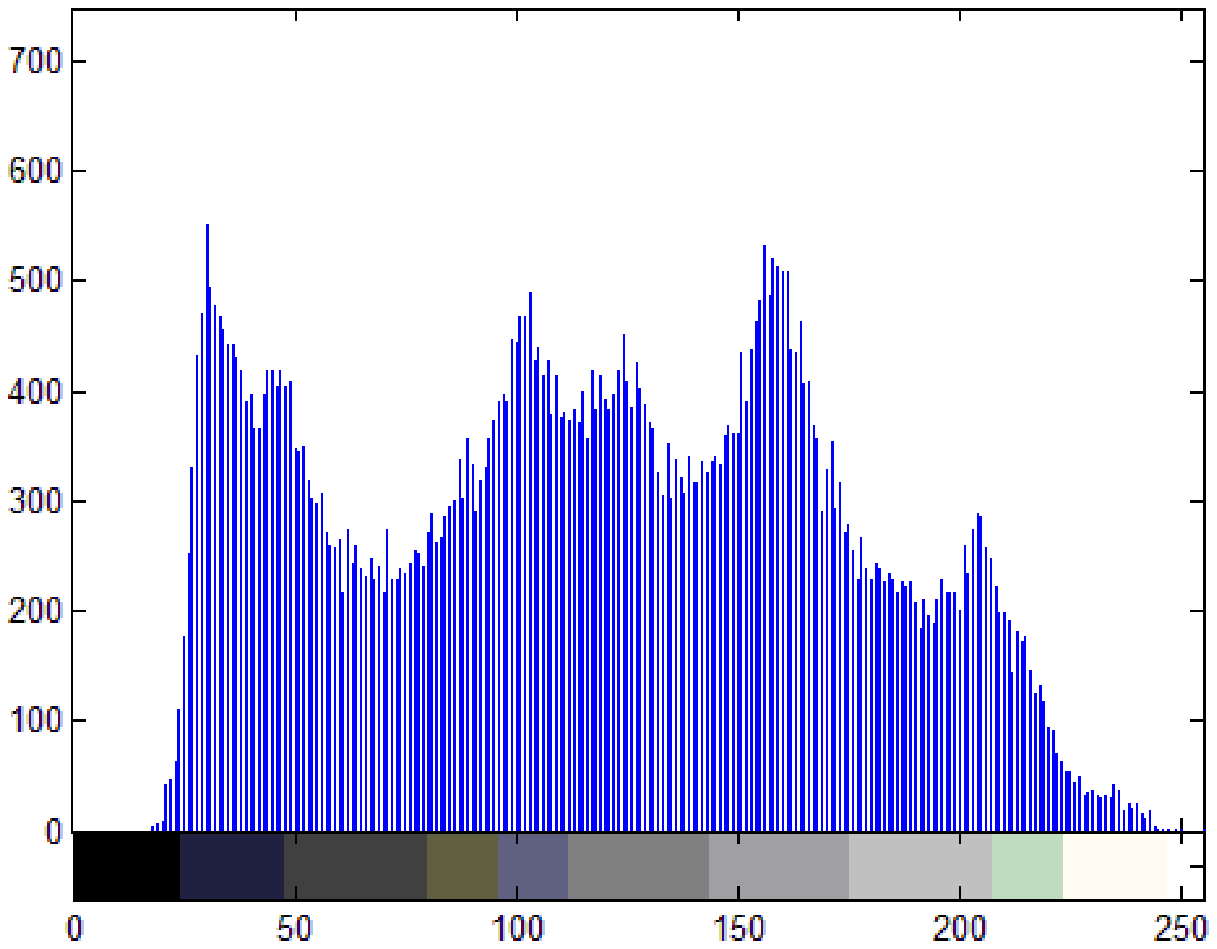}}
\caption{Original Image}
\end{subfigure}
\begin{subfigure}[b]{0.3\textwidth}
\resizebox{7cm}{6cm}{\includegraphics{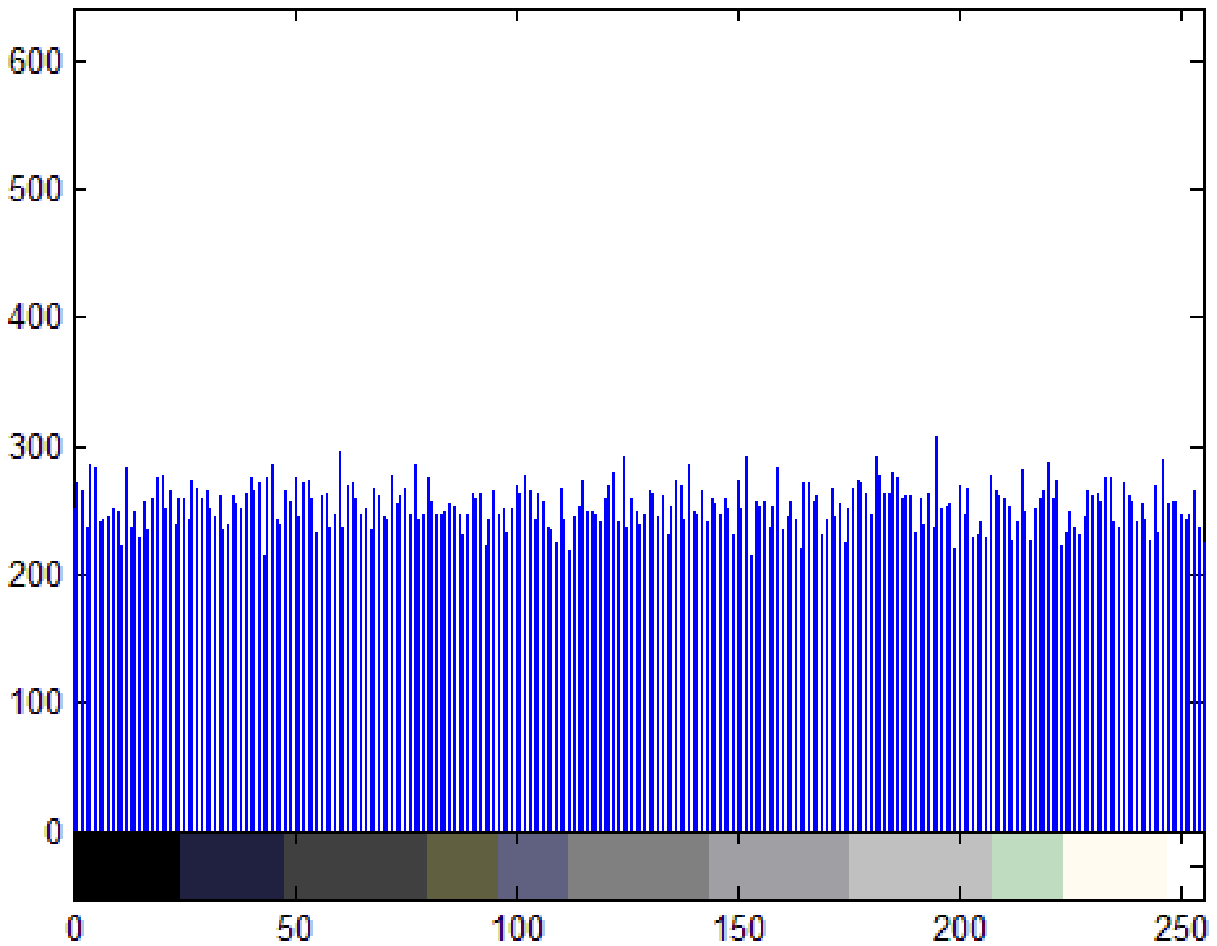}}
\caption{Cipher Image}
\end{subfigure}
\begin{subfigure}[b]{0.3\textwidth}
\resizebox{7cm}{6cm}{\includegraphics{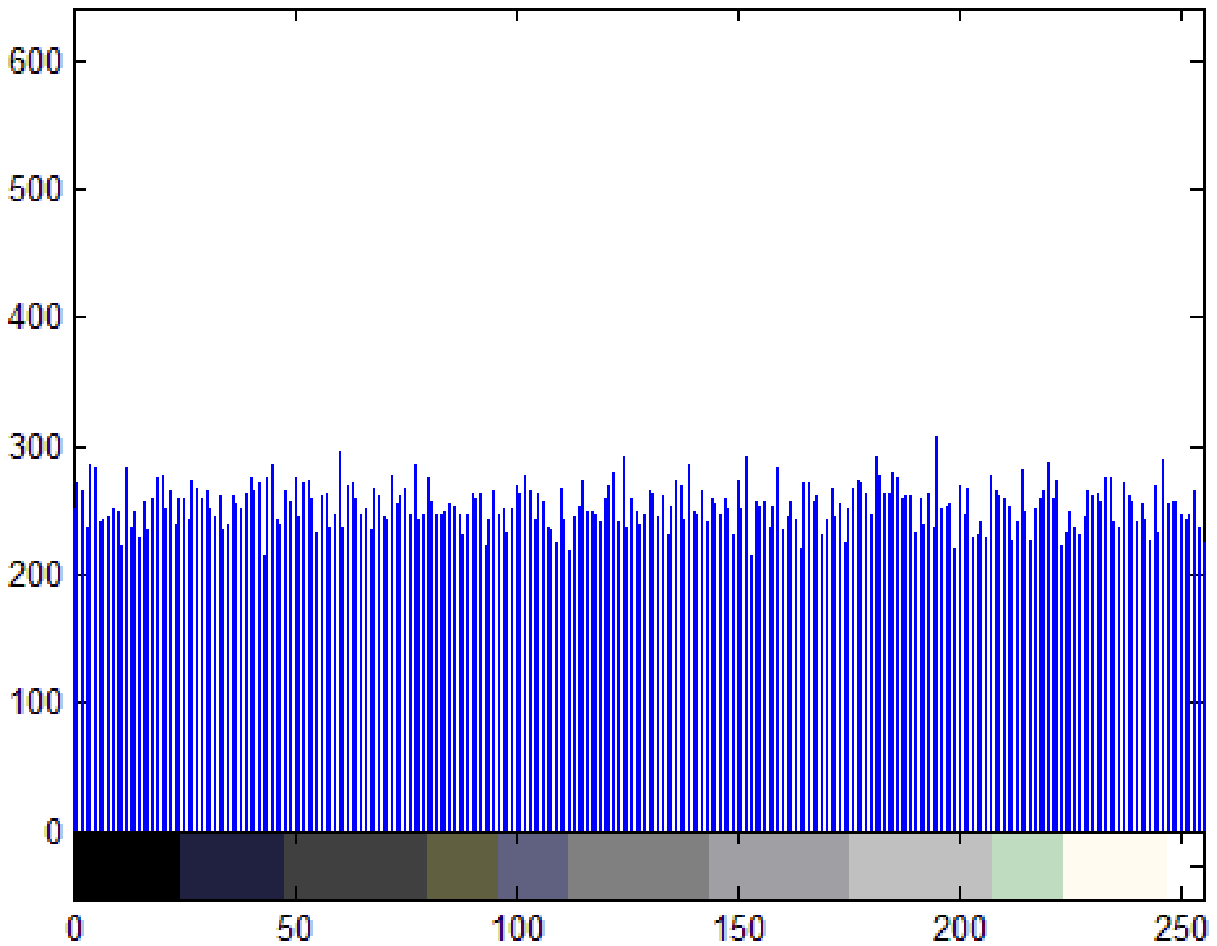}}
\caption{Decrypted Image}
\end{subfigure}
\caption{Image Histograms}
\end{figure}

\subsection{Key Sensitivity Analysis}
An efficient image encryption algorithm should be sensitive to secret key, that is even a minute change in secret key should result into a completely different decrypted image. To check the key sensitivity of proposed algorithm, the initial condition (secret key) $x_0$ is changed a little $10^{-14} ( = \Delta)$, the decrypted image with $x_0 + \Delta $, is shown in Figure 5 which is completely different and unrecognizable. Hence, it can be said that the proposed algorithm has high sensitivity even for a minute change in secret key.
\begin{figure}
\centering
\resizebox{5cm}{5cm}{\includegraphics{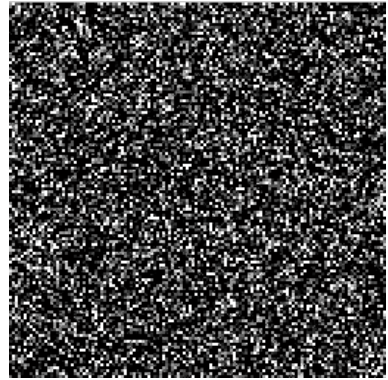}}
\caption{Decrypted Image with incorrect key $x_0 + \Delta$ }
\end{figure}

\subsection{Correlation Analysis}

The correlation coefficient analysis indicates the relationship between pixels in the image. In plain images, there exists a high degree of correlation among adjacent pixels. So, the primary aim of secure image encryption technique is to reduce this correlation [9][10][11]. The correlation coefficient between two adjacent pixels in an image is defined by (9).

\begin{equation}
\rho(x,y) = \frac{\Sigma^{N}_{i=1} \left[ (x_i - E(x_i))(y_i - E(y_i)\right] }{
\sqrt{\Sigma^{N}_{i=1} \left[ x_i - E(x) \right]^2 } \sqrt{\Sigma^{N}_{i=1} \left[ y_i - E(y) \right]^2 }}
\end{equation} 
where $ E(x)=mean(x_i)$, {\it x} and {\it y} are gray values of two adjacent pixels in the image.\\

To determine the encryption quality of proposed technique, the correlation coefficient of 1000 randomly selected pairs of vertically, horizontally and diagonally adjacent pixels is calculated using (9). The correlation coefficients in plain-images and cipher-images are listed in Table 1 and Table 2 respectively. From Table 1 it is observed that the two adjacent pixels in the plain-images are highly correlated to each other. However, the values obtained for cipher-images are close to 0 which shows that the adjacent pixels in the cipher image are highly uncorrelated to each other. This proves that the proposed encryption is successful in reducing the correlation between adjacent pixels. The distribution of two adjacent pixels in the original Barbara image and its encrypted version by the proposed algorithm is shown in Figure 6, 7 and 8. It is clear from these figures and Table 2 that the proposed technique de-correlates the adjacent pixels in cipher image.

\begin{table}
\renewcommand{\baselinestretch}{1}
\caption{Correlation Coefficients of adjacent pixels in plain images}
\begin{small}
\begin{center}
\begin{tabular}{|r|r|r|r|r|} \hline
    Image name & Horizontal & Vertical & Diagonal  \\ \hline

    Lena & 0.9303 & 0.9485 & 0.8588\\  \hline
    Barbara & 0.8499 & 0.8811 & 0.8679\\  \hline
    Baboon & 0.8250 & 0.7944 & 0.7577\\  \hline
    Boat & 0.8799 & 0.9005 & 0.8136\\  \hline
    Peppers & 0.9448 & 0.9532 & 0.9067\\  \hline

\end{tabular}
%\caption{Simulation parameter value setting}
\end{center}
\end{small}
\end{table}

\begin{table}
\renewcommand{\baselinestretch}{1}
\caption{Correlation Coefficients of adjacent pixels in cipher images}
\begin{small}
\begin{center}
\begin{tabular}{|r|r|r|r|r|} \hline
    Image name & Horizontal & Vertical & Diagonal  \\ \hline

    Lena & 0.0252 & 0.0058 & -0.0246\\  \hline
    Barbara & -0.0316 & 0.0243 & 0.0272 \\  \hline
    Baboon & -0.0142 & 0.0034 & -0.0022 \\  \hline
    Boat &0.0135 & 0.0069 & -0.0297 \\  \hline
    Peppers &-0.0045 & -0.0698 & 0.0221\\  \hline

\end{tabular}
%\caption{Simulation parameter value setting}
\end{center}
\end{small}
\end{table}

\begin{figure}
\begin{subfigure}[b]{0.3\textwidth}
\resizebox{6.5cm}{6.5cm}{\includegraphics[width=\textwidth]{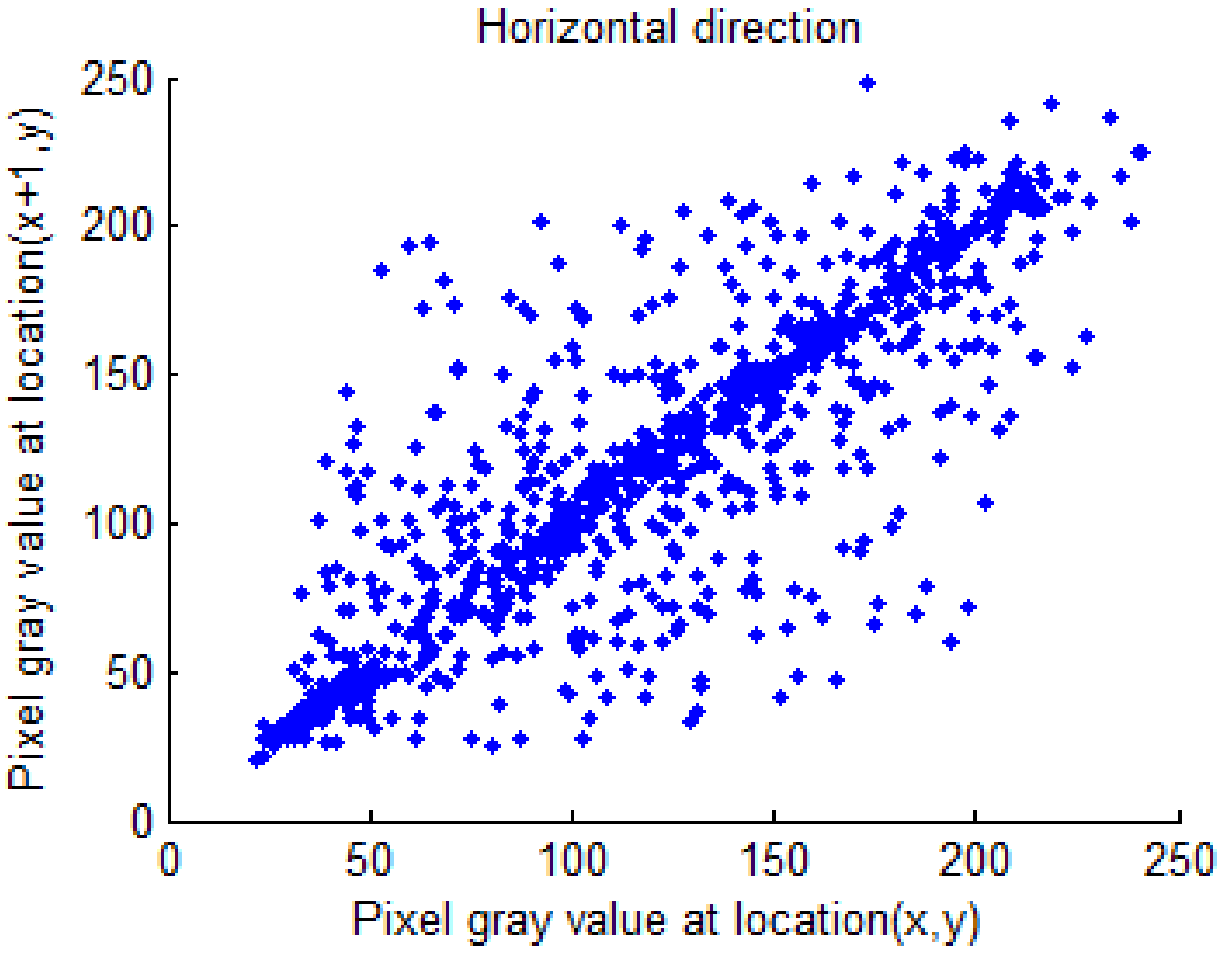}}
\caption{Original Image}
\end{subfigure}
\begin{subfigure}[b]{0.3\textwidth}
\resizebox{7cm}{7cm}{\includegraphics{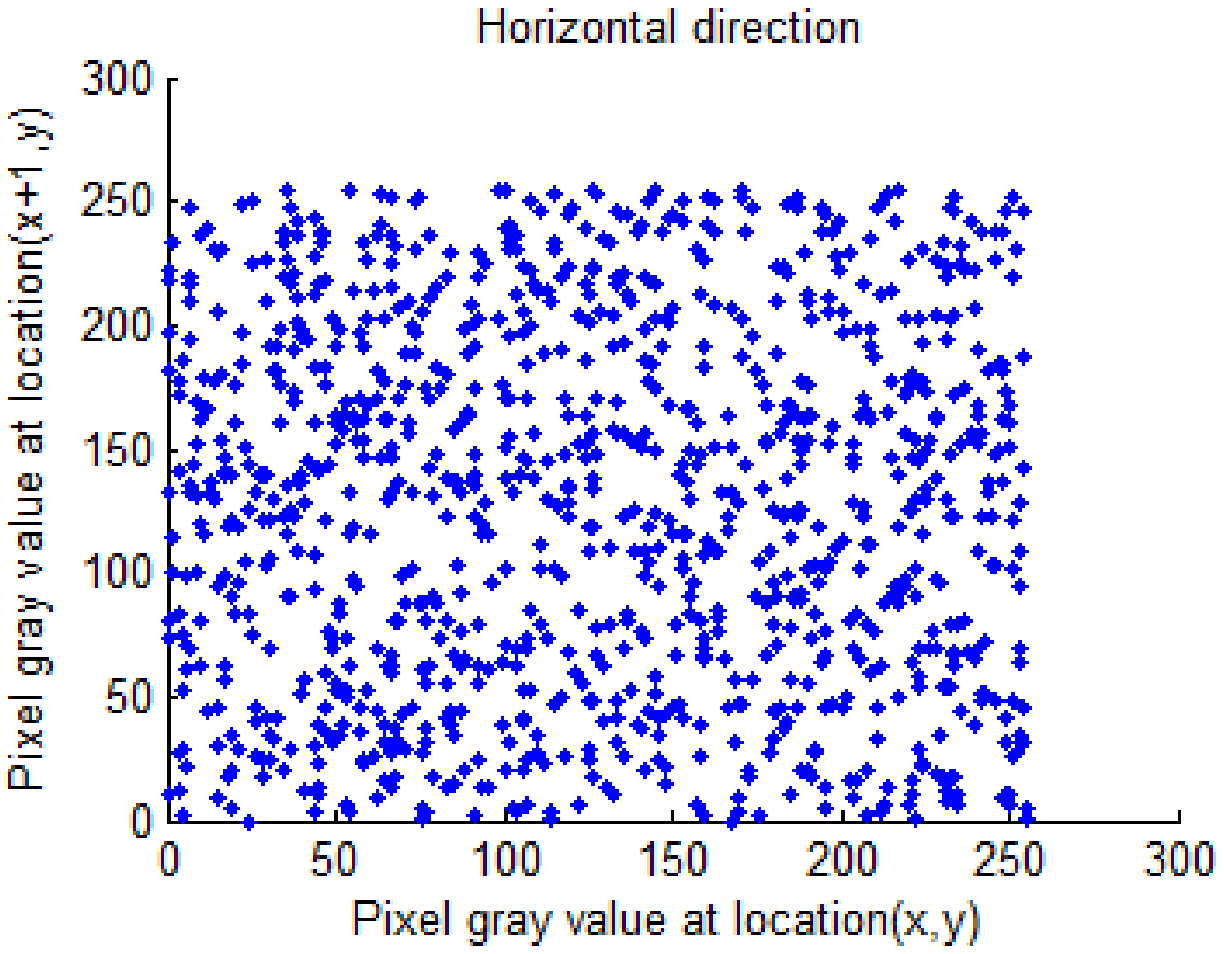}}
\caption{Cipher Image}
\end{subfigure}
\caption{Distribution of adjacent pixels in horizontal direction}
\end{figure}

\begin{figure}
\begin{subfigure}[b]{0.3\textwidth}
\resizebox{6.5cm}{6.5cm}{\includegraphics{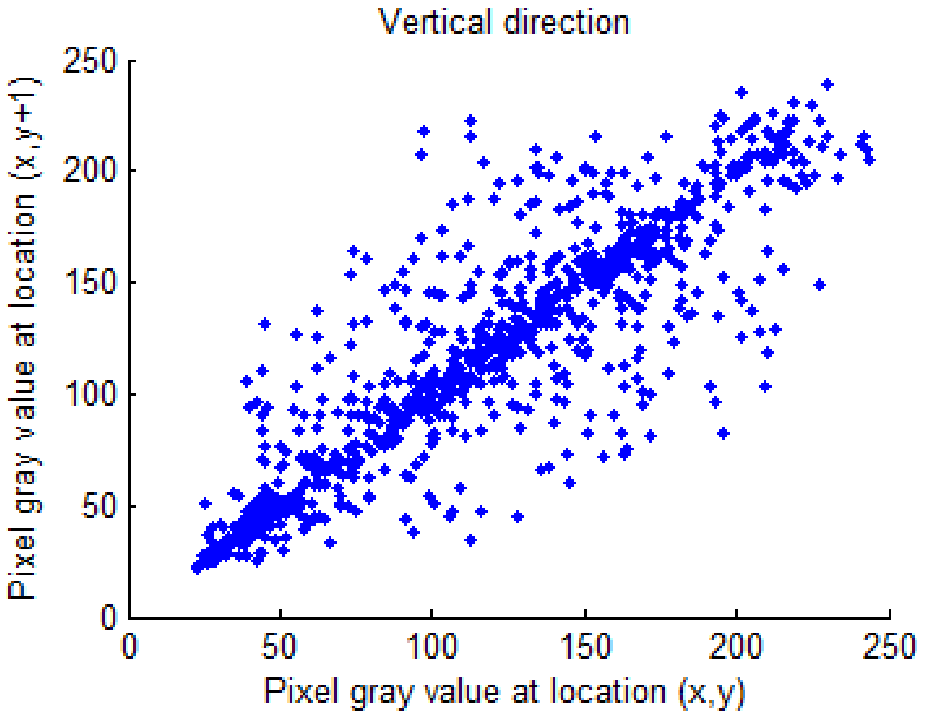}}
\caption{Original Image}
\end{subfigure}
\begin{subfigure}[b]{0.3\textwidth}
\resizebox{7cm}{7cm}{\includegraphics{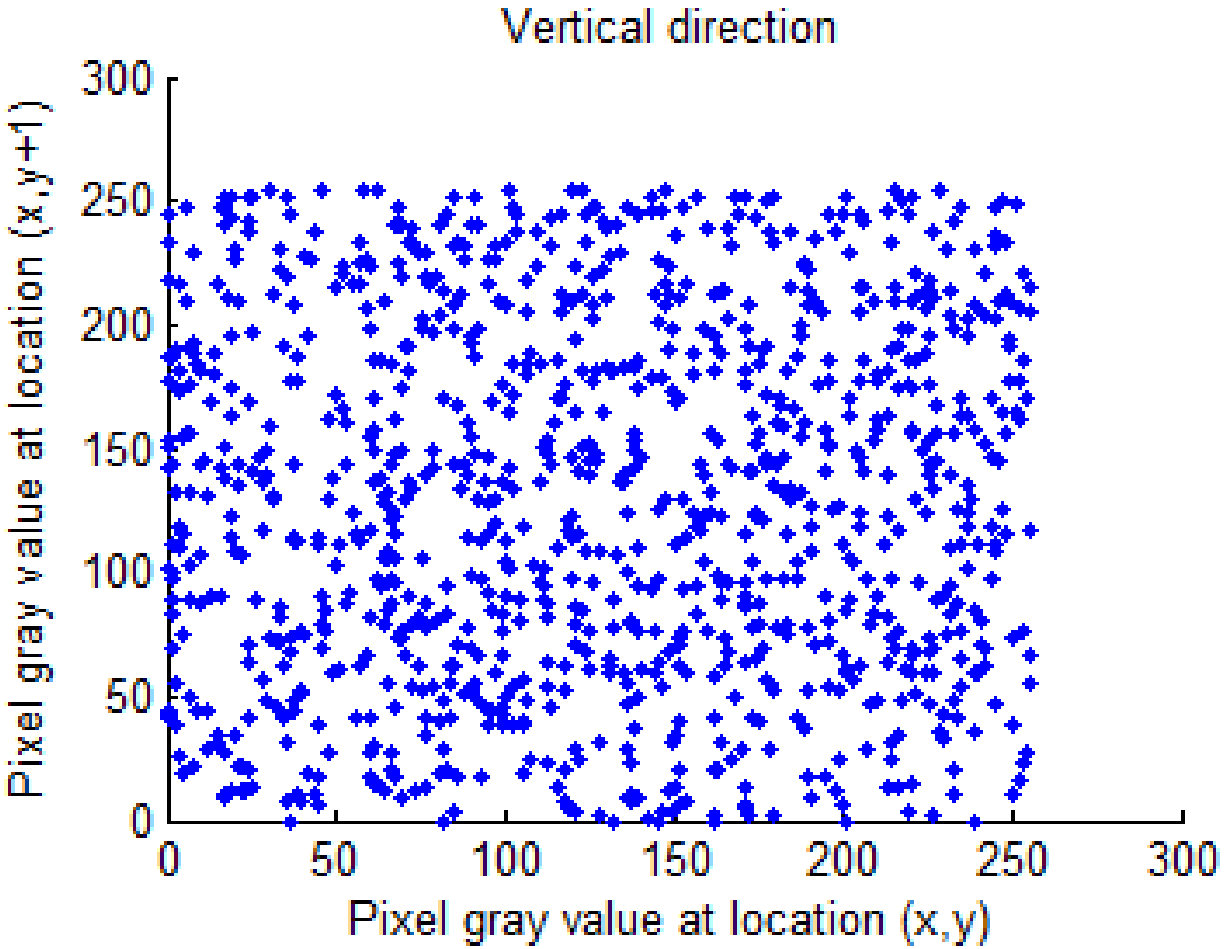}}
\caption{Cipher Image}
\end{subfigure}
\caption{Distribution of adjacent pixels in vertical direction}
\end{figure}

\begin{figure}
\begin{subfigure}[b]{0.3\textwidth}
\resizebox{6.5cm}{6.5cm}{\includegraphics{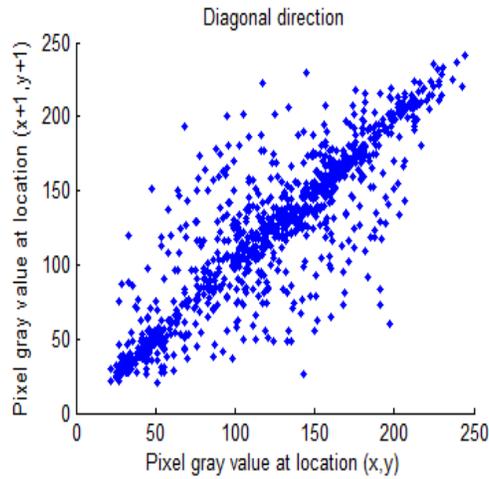}}
\caption{Original Image}
\end{subfigure}
\begin{subfigure}[b]{0.3\textwidth}
\resizebox{6.5cm}{6.5cm}{\includegraphics{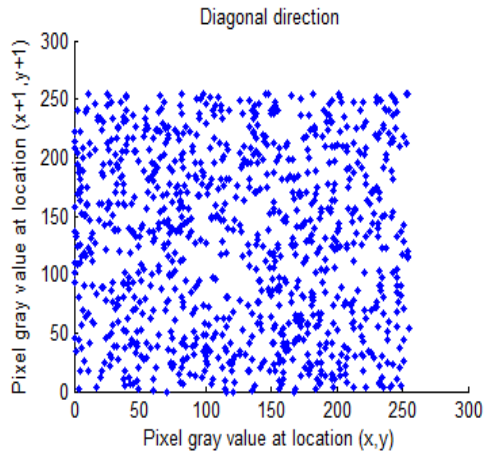}}
\caption{Cipher Image}
\end{subfigure}
\caption{Distribution of adjacent pixels in diagonal direction}
\end{figure}

\subsection{Entropy Analysis}
Entropy is used to measure the uncertainty associated with the random variable. For message source  {\it M}, it can be calculated as defined in (10).
\begin{equation}
H(M) = -\Sigma^{255}_{i=0} p(m_i)log_2(p(m_i))
\end{equation}

where $p(m_i)$ is the probability of symbol $m_i$ and the entropy is expressed in bits. If the message source \textit{M} emits $2^8$ symbols as $M =\lbrace m_0, m_1, . . . , m_{255} \rbrace$ with equal probabilities, then the entropy of \textit{M} is 8, which corresponds to a true random source and represents the ideal value of entropy for \textit{M} [12]. Information entropy of an encrypted image represents the distribution of gray values. Greater value of information entropy indicates the uniformness of distribution of gray value of image [13].\\

A secure system must satisfy a condition on the information entropy that is the encrypted image should not provide any information about the original image. The entropy values obtained for cipher-images of image data set are accumulated in Table 3. It can be concluded from Table 3 that entropy values for cipher images are close to the ideal value 8. This implies that the information leakage in the proposed encryption technique is negligible.

\subsection{Net Pixels Change Rate Analysis}
To test the difference between difference between original plain image \textit{P} and cipher image \textit{C}, net pixels change rate is used. The NPCR scores determine the percentage difference between the plain image and its cipher image [14][15]. For calculation of NPCR value, two-dimensional array \textit{D(i,j)} is defined, having the same size as the images \textit{P} and \textit{C} as, if $P(i,j) \neq C(i,j)$ then $D(i,j) = 1$, otherwise $D(i,j) = 0$. The NPCR is calculated using (11).
\begin{equation}
NPCR = \frac{\Sigma^{M-1}_{i=0} \Sigma^{N-1}_{i=0} D(i,j)}{M \times N}
\end{equation}

This value is calculated for the test image set and tabulated in Table 3. The expected NPCR value for a good encryption technique is 99.6094\%, hence it can be concluded from the table, that NPCR value obtained by the proposed encryption algorithm is significantly good and much closer to expected value.

\begin{table}
\renewcommand{\baselinestretch}{1}
\caption{NPCR and Information Entropy}
\begin{small}
\begin{center}
\begin{tabular}{|r|r|r|r|r|} \hline
    Image name & NPCR & Information Entropy \\ \hline

    Lena & 99.6170 & 7.9966 \\  \hline
    Barbara & 99.6475 & 7.9969 \\  \hline
    Baboon & 99.6368 & 7.9971 \\  \hline
    Boat & 99.6231 & 7.9970 \\  \hline
    Peppers & 99.6154 & 7.9972 \\  \hline

\end{tabular}
%\caption{Simulation parameter value setting}
\end{center}
\end{small}
\end{table}

\subsection{Speed Analysis}
The time is evaluated to assess the speed performance of the proposed algorithm. The algorithm is implemented on Intel core i3 CPU @ 2.1GHz and 4GB RAM memory. The time of encryption and decryption obtained for different images, listed in Table 4, show the speedy performance of the proposed algorithm.

\begin{table}
\renewcommand{\baselinestretch}{1}
\caption{Encryption and Decryption time (in secs)}
\begin{small}
\begin{center}
\begin{tabular}{|r|r|r|r|r|} \hline
    Image name & Enc. time & Dec. time \\ \hline

    Lena & 14.1170 & 14.1966 \\  \hline
    Barbara & 15.1475 & 15.0169 \\  \hline
    Baboon & 14.1248 & 14.1971 \\  \hline
    Boat & 15.1231 & 15.2170 \\  \hline
    Peppers & 14.6154 & 14.5972 \\  \hline

\end{tabular}
%\caption{Simulation parameter value setting}
\end{center}
\end{small}
\end{table}

\section{CONCLUSION}
In this communication, a new chaos based image encryption algorithm is proposed for grayscale images. The algorithm is based on Fisher- Yates shuffle in the wavelet frequency domain. The Fisher-Yates shuffling technique is an efficient approach requiring only time proportional to the number of elements being shuffled. Further, the shuffling effect is increased by incorporating a piece-wise linear chaotic map as source to generate random indices. The random indices are generated from secret key to make shuffling key dependent. Moreover, the chaotic modulation of shuffled approximate coefficients results a good quality of shuffled image. Then, in second phase the shuffled image is self keyed diffused and mixing operation is carried out with the generated keystream sequence and shuffled image. The simulation and experimental analyses proved that the proposed encryption algorithm is secure, efficient, and highly robust towards cryptanalysis.

%\begin{biography}

\noindent{\includegraphics[width=1in,height=1.7in,clip,keepaspectratio]{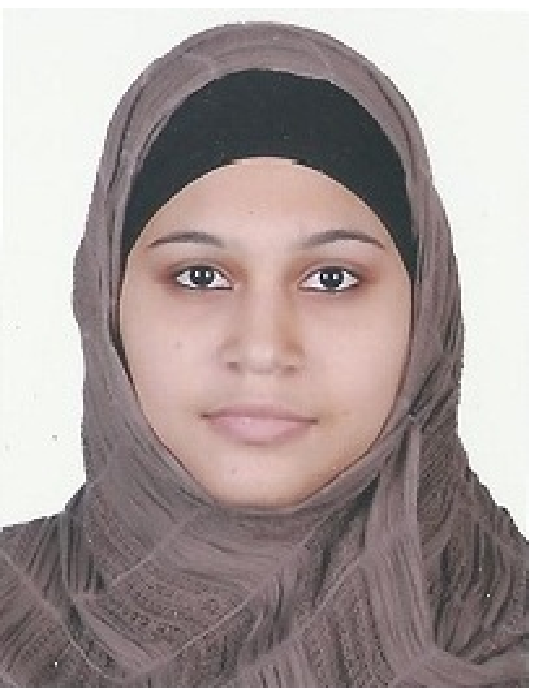}}
\begin{minipage}[b][1in][c]{1.8in}
{\centering{\bf {Swaleha Saeed}} obtained her B. Tech and M. Tech degrees from Department of Computer Engineering, Zakir Hussain College of Engineering and Technology, Aligarh Muslim University, India in 2012 and 2014, respectively.}\\\\
\end{minipage}
She has profound interest in application-end research work and projects that contribute to a social cause. Her areas of research interest are Cryptography, Graphical User Authentication Systems and Image Processing Techniques.\\\\\\

%---------------------------------------------------------------------------------------------------------------------
%\begin{biography}

\noindent{\includegraphics[width=1in,height=1.7in,clip,keepaspectratio]{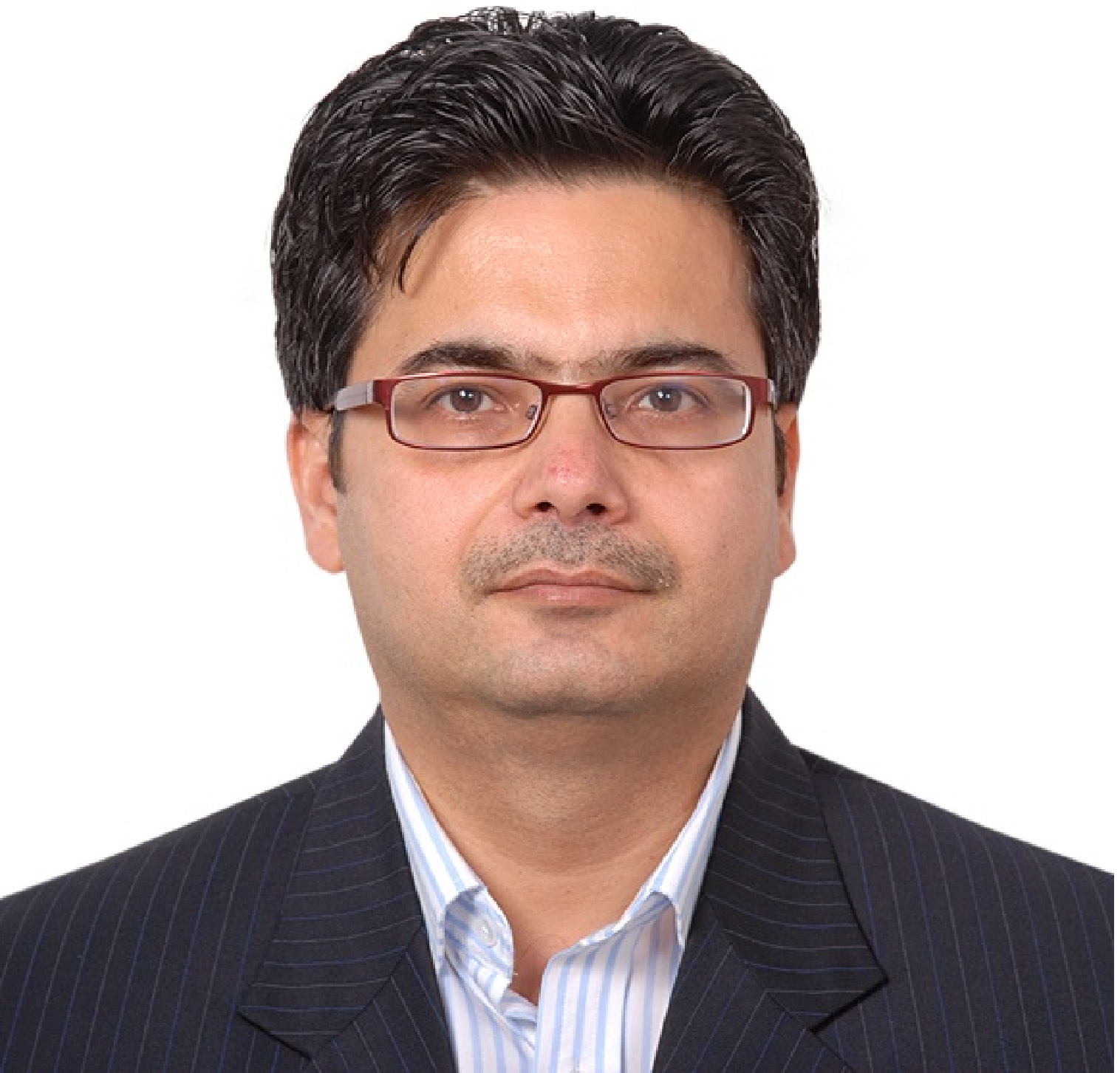}}
\begin{minipage}[b][1in][c]{1.8in}
{\centering{\bf {M Sarosh Umar}} is working as Associate Professor in the Department of Computer Engineering, Zakir Hussain College of Engineering and Technology, Aligarh Muslim University, India. He has teaching/research and indus-}\\\\
\end{minipage}
trial experience of more than 25 years in India as well as abroad. He has various articles, research papers and published patents to his credit. His research interests are User Authentication using Graphical Methods, Computer Security and Software Engineering.\\\\

\noindent{\includegraphics[width=1in,height=1.7in,clip,keepaspectratio]{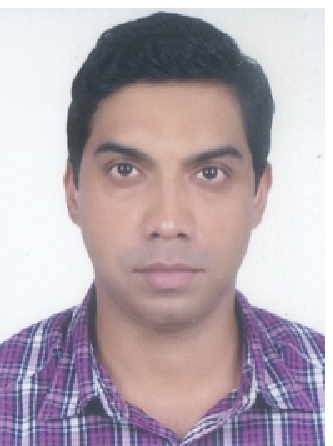}}
\begin{minipage}[b][1in][c]{1.8in}
{\centering{\bf {M Athar Ali}} is currently employed as an Assistant Professor in the Department of
Computer Engineering, Zakir Hussain College of Engineering and Technology, Aligarh Muslim University (AMU), India. Dr. Ali earned his PhD from Loughborough}\\\\\\
\end{minipage}
University, England, United Kingdom and his masters and graduate degrees from AMU. His areas of interest include but are not limited to Image and Video coding/processing, Information security and Cryptography.\\\\\\\\

\noindent{\includegraphics[width=1in,height=1.7in,clip,keepaspectratio]{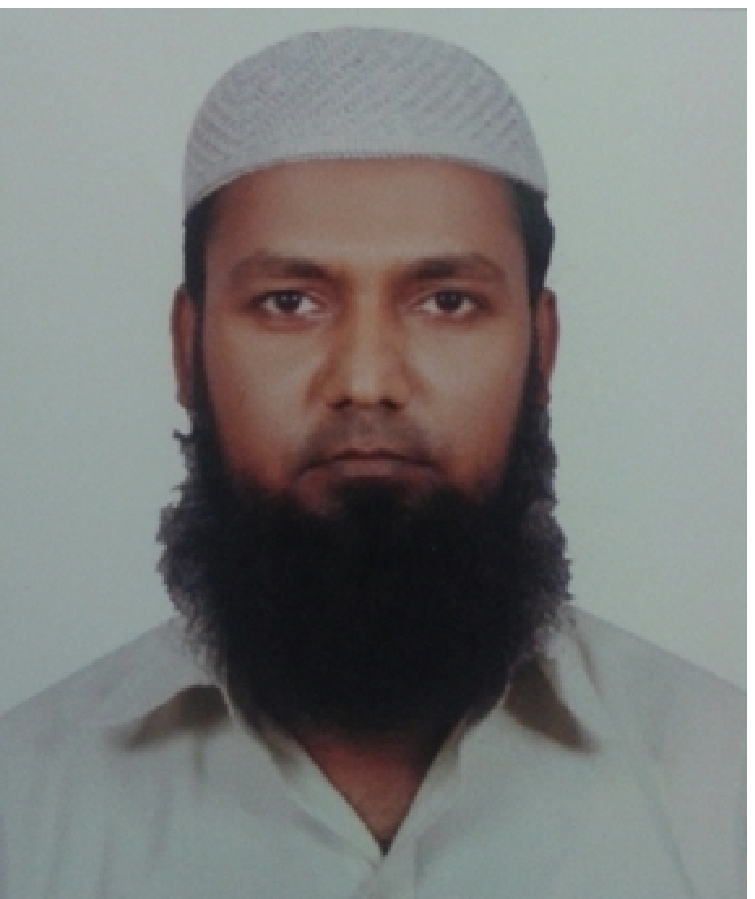}}
\begin{minipage}[b][1in][c]{1.8in}
{\centering{\bf {Musheer Ahmad}} received his B. Tech and M. Tech degrees from Department of
Computer Engineering, Zakir Hussain College of Engineering and Technology, Aligarh Muslim University, India in 2004 and 2008, respectively.}\\\\
\end{minipage}
He is with Department of Computer Engineering, Jamia Millia Islamia, New Delhi, India, as Assistant Professor. He has published about 30 research papers in refereed academic journals and international conference proceedings. His areas of research interest include Multimedia Security, Chaos-based Cryptography, Cryptanalysis and Image Processing.

\end{document}